\documentclass[fleqn,12pt,twoside]{article}
\usepackage{espcrc1}
\readRCS
$Id: espcrc1.tex,v 1.2 2004/02/24 11:22:11 spepping Exp $
\usepackage{graphicx}
\usepackage[figuresright]{rotating}

\newcommand{\AmS}{{\protect\the\textfont2
  A\kern-.1667em\lower.5ex\hbox{M}\kern-.125emS}}

\title{{QQ-onia} package: a numerical solution to the
  Schr\"{o}dinger radial equation for heavy quarkonium}

\author{Juan-Luis Domenech-Garret \footnote{Corresponding author E-mail: domenech@macs.udl.es} \address {Departamento MACS, F\'{\i}sica Aplicada.
Universitat de Lleida.\\
  Av. Alcalde Rovira Roure 191, Bldg-3, E-25198, Lleida (Spain)}\\
Miguel-Angel Sanchis-Lozano \footnote{E-mail:Miguel.Angel.Sanchis@uv.es}
\address {Instituto de F\'{\i}sica
Corpuscular (IFIC) and Departamento de F\'{\i}sica Te\'orica,\\
Centro Mixto Universitat de Val\`encia-CSIC \\
Dr. Moliner 50, E-46100 Burjassot, Valencia (Spain)}}

\begin{document}

\maketitle

\begin{abstract}
 This paper presents the basics of the \emph{QQ-onia} package, a
 software based upon the Numerov $\textit{0}(h^6)$ method which can
 be used to solve the Schr\"{o}dinger radial equation  using a
 suitable potential $V(r)$ for the heavy quarkonium system. This
 package also allows the analysis of relevant properties of those
 resonances such as the square of the wave functions at the origin, their
 corresponding derivatives for $l \neq 0$ states, typical heavy-quark
 velocities, and mean square radii. Besides, it includes a tool to 
 analyze the  spin dependent contributions to the heavy quarkonia  
 spectrum, providing the splitting of  $n^3S_1-n^1S_0$, as well as  
 the $n^3P_J-n^1P_1$ energy levels. Finally a simple software implemented
in \emph{QQ-onia} to compute E1 transition rates is presented.   
\end{abstract}

\begin{description}

\item[]PROGRAM SUMMARY

\item[]Program title: QQ-ONIA PACKAGE\footnote{Program Author and Copyright: Juan-Luis Domenech-Garret}.

\item[]Manuscript title: QQ-onia package: a numerical solution to the
  Schr\"{o}dinger radial equation for heavy quarkonium.

\item[]Manuscript Authors: Juan-Luis Domenech-Garret;  Miguel-Angel Sanchis-Lozano.

\item[]Programming language: PAW (Physics Analysis Workstation).

\item[]Operating system(s) for which the program has been designed: Windows-XX and  Unix
(Linux).

\item[]Keywords: Heavy Quarkonium potential; Wave function at the origin.

\item[]PACS: 14.40.-n; 12.39.-x;  14.40.Gx

\item[]Nature of the problem: Software  to solve the Schr\"{o}dinger radial equation  using a
 suitable potential $V(r)$ for the heavy quarkonium system, allowing to perform spectroscopy. It
 also allows the analysis of relevant quantities of those
 resonances such as the square of the wave functions at the origin, their
 corresponding derivatives for $l \neq 0$ states, typical heavy-quark
 velocities, and mean square radii. The package is a (user-friendly) multipurpose tool
 for dealing with different heavy quarkonium systems, providing  a way
 to study the influence of a given potential on a series of relevant
 physical quantities, by either  varying parameterized values of a well-known
 potential form, or by including new terms.

\item[]Solution method: Based upon the Numerov $\textit{0}(h^6)$ Method,
we perform a matching procedure
to  the reduced wave function at the cut point. We also perform a
normalization technique  for  for these wave functions taking into
account the different domains when we use a Numerov backward-forward
technique. In the case of $l \geq 2$ we  present a way to find
the  corresponding  derivatives  at the origin by only calculating
the reduced  radial wave function and  first
derivative. When estimating the heavy  quark velocity, we
 introduce an additional way to compute this quantity from  the virial theorem.
 The calculated reduced wave functions and radial wave functions at 
the origin are later used to obtain the heavy quarkonia $nL$ splitting and 
$E1$ transition rates.  

\item[]Additional comments: Using Windows, to optimize the edition of the files, please, open it with MFC-WORDPAD.

\item[]Running time: It depends on the choice of the $r$ range, and the number of energy steps.

\end{description}

\newpage

\tableofcontents

\newpage

\section{Introduction}
Since the discovery of the charmonium and bottomonium families,
much efforts have been spent over 30 years trying to understand the nature
 of heavy quarkonium, (as a summary see for example
 \cite{Brambilla}); in the meantime various numerical tools have been
 created in an almost $\lq\lq$ad hoc" fashion with the aim of
 extracting important information about their properties, such as
 their masses and partial widths.\\

 This package was developed with the aim of providing a multipurpose 
(user-friendly)  tool
 for dealing with different heavy quarkonium systems, providing  a way
 to study the influence of a given potential on a series of relevant
 physical quantities, by either  varying parameterized values of a well-known
 potential form, or by including new terms.\\

The \emph{QQ-onia} package handles the heavy quarkonium system
within a  non relativistic framework, solving the Schr\"{o}dinger
radial  equation (\emph{SRE}) with an appropriate potential for
heavy  quarks/antiquarks.
 The basic reason for this choice derives
 from the Quark Potential Model \cite{Rosner11},
 which establishes a low value for the expected square velocity of the quark for these
 heavy resonances ($v^2\sim 0.1$ for bottomonia  and $v^2\sim 0.3$ for charmonia).
  Besides, there is also another reason from a dual ultra-relativistic picture
  to provide a non relativistic treatment for heavy quarkonium \cite{schob}.
  These low velocities $v^2<<1$  were also responsible for the success of the
   Non Relativistic QCD (NRQCD) (\cite{BBLep}, \cite{Eiras}), a rigorous effective
  theory for strong interactions deriving from first principles.\\

The paper is organized as follows: the first section explains the basics of the heavy quarkonium
Non Relativistic potential and its spin-depedent part, according with the Breit-Fermi Hamiltonian. 
Later we explain  the   underlying foundations to our way of solving the
 Schr\"{o}dinger equation. Although the arguments presented are
 already  well-known, we consider this introduction necessary to
 facilitate the understanding of the rest of the article. We then
 look into the specific details about the calculation of the wave functions of
 heavy quarkonia in our code. 

We initially focus on  a
 procedure for matching the reduced wave function at the cut
 point and its normalization when we use a Numerov backward-forward technique. We
 also explain  how to extract useful quantities in order to analyze the heavy
 quarkonium system such as  mean square radius, and heavy  quark velocity
 (introducing an additional way to calculate this quantity from  the virial theorem).
  For $l=0$ states we also focus on square of the wave functions at the origin, and, for $l \neq 0$ states,
 we examine their  corresponding  derivatives; in the case of $l \geq 2$ we
 present a way to find  these values by only calculating their corresponding  reduced
 radial wave function and  first derivative. It follows an explanation on the spin-dependent 
 terms in the bottomonia case and, finally, we focus on  the calculation  of the  E1 transition rates.\\

   The software will be explained in more detail in
  the second part of the paper, where we also present several
 results obtained using the programme to illustrate the
procedure and foreseen precision.\\
  
 Several potentials for the heavy quarkonium system are given as examples. 
Our first (main) choice is the well-known funnel or Cornell potential,
i.e. a Coulomb plus Lineal static potential (CpL) \cite{eitchten2}, 
 
\[
V(r)= \sigma r - \frac{C_F \alpha_s \hbar}{r}
\] 
with string tension $\sigma$ and strong coupling
  constant $\alpha_s$; $C_F$ is a colour factor.\\
 
Nevertheless, other static and non-static potentials 
will also be used in this paper in order to illustrate some results. 
Note that we will focus on the equal masses. Thus one can write
for bottomonium $m_Q=m_{\bar Q}= m_b\equiv m$ with  
reduced mass $\mu = m_Q/2$.
   
\section{Physical Bases of the package}

\subsection{Heavy quarkonium potential and spin-dependent terms}\

If $V_{NR}$ stands for the Non Relativistic potential, one 
can split it in two terms
 consisting of a vector ($V$) and  a scalar ($S$) contribution \cite{schob2,schob3}
 
\begin{equation}
 V_{NR}(r)\ =\  V_{V}(r)\ +\  V_{S}(r)
\end{equation}

 In our example with a funnel type  potential (being  $k \equiv C_F \alpha_s \hbar$) 
\[
V_{V}(r)=\ -k/r\ ;\  V_{S}(r)=\ \sigma r 
\]

In accordance with literature (e.g. \cite{schob2}), additional terms
have been included in 
the potential $V_{NR}$ to take into the account 
the spin orbital and the spin-spin interactions,
causing the splitting of the different mass levels. 
The additional potential reads \cite{schob3}

\begin{equation}
V(r)_{spin-dependent}\ =\  V_{LS}\ +\  V_{SS}\ +\  V_{T}
\end{equation}

where $V_{LS}$, $V_{SS}$, and $V_{T}$ are the spin-orbit, the spin-spin, and the 
 tensor terms, respectively. The spin-orbit term, 
in the equal quark masses case, is 

\begin{equation}
V_{LS}(r)= \frac{(\bf{L\cdot S})}{2\ m^2\ r}\  \biggl[\ 3\ \frac{d}{dr}V_{V}(r)\ -\ \frac{d}{dr}V_{S}(r)\ \biggr]
\end{equation}
where $\bf{L}$ is the relative angular momentum of the 
constituents (1 and 2),and $\bf{S}$ is the total spin of the bound
 state, $\bf{S \equiv S_1 + S_2}$ ( with $\bf{J\equiv L+S}$ ); 
$\langle (\bf{L\cdot S}) \rangle$  
for different $j$ and $l$ values is shown in Table 1.  

\begin{table}[htb]
\caption{$\langle (\bf{L\cdot S}) \rangle$ 
coefficients ($\langle \bf{L\cdot S} \rangle $ 
$= 0$\  if\  $l=0$\ or\  $S=0$).}
\label{table:9}
\newcommand{\m}{\hphantom{$-$}}
\newcommand{\cc}[1]{\multicolumn{1}{c}{#1}}
\renewcommand{\tabcolsep}{2pc} 
\renewcommand{\arraystretch}{1.2} 
\begin{tabular}{@{}|l|l|l|l|}
\hline
 $j$ value & $(l+1)$ & $  l  $ & $(l-1)$ \\
\hline
$\langle \bf{L\cdot S} \rangle $  & $l$ & $-1$ & $-(l+1)$\\
\hline
\end{tabular}\\[2pt]
\end{table}

The spin-spin term can be written as

\begin{equation}
V_{SS}(r)= \frac{2\ (\bf{S_1\cdot S_2})}{3\ m^2\ r}\  \biggl[\ \Delta(\ V_{V}(r)\ )\ \ \biggr]
\end{equation}
where the $\langle  \bf{S_1\cdot S_2} \rangle $ coefficients
take the values:\\
 
$(-3/4)$ for the spin-singlet case ($S=0$), and $(+1/4)$ for 
the spin-triplet case ($S=1$).\\

The tensor term can be written as

\begin{equation}
V_{T}(r)= \frac{1}{12\ m^2} (S_{12})\  \biggl[\ \frac{1}{r}\ \frac{d}{dr}V_{V}(r)\ -\ \frac{d^2}{dr^2}V_{V}(r)\ \biggr]
\end{equation}
where $(S_{12})$ is the spin-dependent factor (for $l\neq 0$ and $S=1$),
shown in Table 2 for different $j$ and $l$ values.\\
  
\begin{table}[htb]
\caption{ Spin-dependent $S_{12}$ factor. ( $\langle  S_{12} \rangle  
= 0$  if $l=0$ or $S=0$ )}
\label{table:10}
\newcommand{\m}{\hphantom{$-$}}
\newcommand{\cc}[1]{\multicolumn{1}{c}{#1}}
\renewcommand{\tabcolsep}{2pc} 
\renewcommand{\arraystretch}{1.6} 
\begin{tabular}{@{}|l|l|l|l|}
\hline
 $j$ value & $(l+1)$ & $  l  $ & $(l-1)$ \\
\hline
$\langle  S_{12} \rangle $ & $-\ 2\ l /(2l+3)$ & $2$ & $-\ 2\ (l+1)/(2l-1)$ \\
\hline
\end{tabular}\\[2pt]
\end{table}

\subsection{Schr\"{o}dinger radial equation}\

 Basically, our code has to solve the well known Schr\"{o}dinger radial 
equation (\emph{SRE}):

\begin{equation}
\frac{d^2}{dr^2}u_l(r)+ \frac{2\mu}{\hbar^2} \biggl[ [E-V(r)]-
 \frac {\hbar^2 l(l+1)}{2\mu r^2} \biggr]u_l(r) =0
\end{equation}
  where $\Psi(r,\theta,\phi)=R_{nl}(r)\  Y_{lm_l}(\theta,\phi)$ is the
  complete wave function, $r$ stands for the \emph{relative} radial coordinate,
  and\ $u_l(r)\equiv r\  R_{nl}(r)$\  is the reduced radial wave
  function.\\

 With respect to the boundary conditions, a regular solution near the
  origin for $u_{l}(r)$ could be \cite{gasio}

\begin{equation}
u(r \rightarrow 0) \rightarrow r^{l+1}
\end{equation}

 Since asymptotically $u(r \rightarrow \infty) \rightarrow 0 $ we can take:

\begin{equation}
u(r \rightarrow \infty) \rightarrow \exp \biggl[ - \frac{\sqrt{2\mu
|E|}}{\hbar} r \biggr]
\end{equation}

 where $|E|$, as later will be seen, is an educated guess about the
  energy eigenvalue.\\

 The normalization condition reads

\begin{equation}
\int_0^\infty\,dr |u_l(r)|^2=\int_0^\infty\,dr |R_l(r)|^2 r^2 =1
\end{equation}

\section{Numerical solution of {\em SRE}}\

 The {\em SRE} can be written as

\begin{equation}
\frac{d^2u(r)}{dr^2}\ +\  k(r)\  u(r) =s(r)
\end{equation}

Here (setting $l = 0$), $k(r)\equiv \frac{2\mu}{\hbar^2}\ [E-V(r)]$ is
 the kernel of the  equation, and $s(r)\equiv 0$.\\

We can integrate these  equations by means of the  Numerov
 Algorithm \cite{koonin} as follows: \\

 First we split the $r$ range into N points according to
 $r_{n}=r_{n-1}+h$ (where h is the step); then we write the wave function
 $u_n\equiv u(r_{n})=u(r_{n-1}+h)$,  and $k_n\equiv k(r_n)=k(r_{n-1}+h)$.\\

Expanding $u(r)$ around $r_n$:\\

\begin{equation}
 u_{n+1} \equiv u(r_{n}+ h)= u(r_{n})+ h u'(r_n) +
\frac{h^2}{2}u^{''}(r_n) + \frac{h^3}{6}u^{'''}(r_n)+
\frac{h^4}{24}u^{(iv)}(r_n)+ \textit{0}(h^5)\nonumber
\end{equation}

\begin{equation}
u_{n-1} \equiv u(r_{n} - h)= u(r_{n})- h u'(r_n)+
\frac{h^2}{2}u^{''}(r_n) - \frac{h^3}{6}u^{'''}(r_n)+
\frac{h^4}{24}u^{(iv)}(r_n)+ \textit{0}(h^5)\nonumber
\end{equation}

Then approximating the second
 derivative by the three-point difference formula, and using it within the
 second-order  differential equation we get the following recursive formulas,
 with a local error \textit{0}($h^6$):\\

\noindent
{\bf a)} Forward recursive relation
\begin{equation}
u_n=\frac{2(1-\frac{5h^2}{12} k_{n-1})\ u_{n-1}\ -\ (1+\frac{h^2}{12}
k_{n-2})\ u_{n-2}}{(1\ +\ \frac{h^2}{12} k_n)}
\end{equation}

\noindent
{\bf b)} Backward recursive relation
\begin{equation}
u_{n-1}=\frac{2(1-\frac{5h^2}{12} k_{n})\ u_{n}\ -\ (1+\frac{h^2}{12}
k_{n+1})\ u_{n+1}}{(1\ +\ \frac{h^2}{12} k_{n-1})}
\end{equation}

Therefore, when we calculate our wave function  using the
 backward-forward technique, we should note that the recursive formulas imply
 having knowledge of two initial values
 for each direction.\\

It is also necessary to know the first derivative at
 the appropriate order. Following the above expansions, we then get:

\begin{equation}
u^{'}_{n}= \frac{1}{2h} \biggl[(1+ \frac{h^2}{6} k_{n+1})\ u_{n+1}\ -\  (1+ \frac{h^2}{6}
k_{n-1})\ u_{n-1} \biggr] + \textit{0}(h^4)
\end{equation}

\section{Wave function and normalization}

For the sake of simplicity, let us first focus
  on $l=0$ states. The \emph{SRE} reduces to

\begin{equation}
\frac{d^2u(r)}{dr^2}+ \frac{2\mu}{\hbar^2}\ [E-V(r)]\  u(r) = 0
\end{equation}

To illustrate this point we choose an harmonic oscillator potential
 $V(r) \equiv \beta r^2$, with $\beta$ a constant. We then start,
  for instance, by using a forward calculation 
(with an appropriate energy eigenvalue).\\

 Since we are dealing with bound states, we find
 eigenfunctions at the classically allowed region with $E>V(r)$
  and classically forbidden region where $E<V(r)$, they are
  separated at a turning point, $r_c$,
  which can be estimated from the equality $E=V(r_c)$.
  If we perform a forward calculation, its asymptotic solution at
 the forbidden region may behave either as $\sim e^{\pm \alpha r^2}$,
 where  the positive value is  non-physical.
 Thus, we have an admixture of those solutions and then, with successive
 iterations, the integration would be numerically
 unstable due to the dominance of the exponentially growing solution.
 As a general rule \cite{koonin},
 integration \emph{into} a classical forbidden region tends to be
 inaccurate.\\

 Hence, for a given energy eigenvalue, we consider a
 calculation using both forward
  and backward solutions: from the allowed towards the forbidden region,
  with $u_{out}(r)$ (outwards) eigenfunctions
  and from the forbidden towards the allowed
  region with $u_{in}(r)$ (inwards) eigenfunctions .\\

Let us note here that, to avoid numerical
 overflows in the forward calculation, we do not usually
  start with $u(r=0)$: once
 included the centrifugal barrier term, the $1/r^{2}$ piece
 would originate an overflow at $r=0$.\\

\subsection{Bound state energy}\

Since both $u_{out}(r)$ and  $u_{in}(r)$ satisfy an homogeneous
 equation, their normalization can always be chosen so that they are set to be
 equal at the $r_c$ point. An energy eigenvalue is then signaled by
 the equality of derivatives at this point \cite{koonin}. At the matching
 point the eigenfunctions $u_{out}(r)$ and  $u_{in}(r)$ and first derivatives
 $u'_{out}(r)$ and  $u'_{in}(r)$ must all satisfy the continuity conditions:

\begin{equation}
\biggl( u_{out} \biggr)_{r_{c}}= \biggl( u_{in} \biggr)_{r_{c}} \\
\biggl( u'_{out} \biggr)_{r_{c}}= \biggl( u'_{in} \biggr)_{r_{c}}
\end{equation}

thus, we can write the corresponding condition for
 the logarithmic derivative at $r_c$ as

\begin{equation}
\biggl[\frac{u'_{out}}{u_{out}}\biggr]_{r_{c}}=\biggl[\frac{u'_{in}}{u_{in}}\biggr]_{r_{c}}
\end{equation}

and then we can define a $G(E)$ function at $r_c$
 \emph{whose  zeros correspond to the energy eigenvalues} as

\begin{equation}
G(E) \equiv \biggl[\frac{u'_{out}}{u_{out}}\biggr]_{r_{c}} -
\biggl[\frac{u'_{in}}{u_{in}}\biggr]_{r_{c}}
\end{equation}

Therefore we proceed numerically  in the following way: we set a
 trial energy range splitting this $E$ range into N points,
 according to $E_{n}=E_{n-1}+\Delta_E$, where $\Delta_E$ is the
 energy step. For each $E_{n}$ we calculate their eigenfunctions
 $u_{out}$ and  $u_{in}$ at the $r_c$ point; and we build the $G(E)$
 function here, looking for a \emph{change of sign} in it (which
 implies a zero cross). Once we find it, we perform a fine tuning
 closing the energy range  until the required tolerance.\\

\subsection{Matching eigenfunctions at the $r_c$ point}\

When we find the energy eigenvalue,  the calculated inwards and outwards
 eigenfunctions will tend not to match at the $r_c$  point.
 However we can look for a strategy to solve this problem:\\

Denoting the outwards and inwards functions \emph{directly} obtained from
 the recursive formulas as $\Phi(r)$
 and $I(r)$, respectively, the physical $u_{out}(r)$ and
 $u_{in}(r)$ eigenfunctions  can be rewritten as

\begin{equation}
u_{out}(r)= A \Phi (r) \\
u_{in}(r)= B I (r)
\end{equation}

 $A$ and $B$ are  constants. Their respective derivatives are

\begin{equation}
u'_{out}(r)= A \Phi'(r) \\
u'_{in}(r)= B I'(r)
\end{equation}

By substituting eqs. $(20)$ and $(21)$ into eq.(17):

\begin{equation}
\biggl( A \Phi \biggr)_{r_{c}}= \biggl( B I \biggr)_{r_{c}} \\
\biggl( A \Phi' \biggr)_{r_{c}}= \biggl( B I' \biggr)_{r_{c}}
\end{equation}

and performing the difference, we get

\begin{equation}
A = \biggl[ \frac{I - I'}{\Phi - \Phi'}\biggr]_{r_{c}} B \equiv
f_{c}\  B
\end{equation}

where $f_{c}$ will be a scaling factor to be applied to $u_{out}(r)$.
Therefore

\begin{equation}
u_{out}(r)= B\  f_{c}\  \Phi (r) \\
u_{in}(r)= B\  I (r)
\end{equation}

and $B$ is a global factor that must be taken into account in the normalization process.

\subsection{Normalization}\

Once the energy eigenvalue has been determined, we first insert it
 into the kernel, $k(r)$, thereby generating
   their corresponding $\Phi(r)$ and  $I(r)$ functions;
subsequently we calculate the $f_c$ factor. 
To find the remaining $B$ factor, and therefore find the $u_{out}$ and
   $u_{in}$ eigenfunctions,  we  use the  normalization condition

\begin{equation}
\int_0^{r_{max}}\,dr |u_l(r)|^2=1
\end{equation}
where, following the asymptotic requirement
$u(r\rightarrow \infty) \rightarrow  0$, 
taking $r_{max}$ as a cutoff value.\\

 By separating the  $u_{out}(r)$ and  $u_{in}(r)$ domains in the
   above integral, we can write

\begin{equation}
\int_0^{r_{max}}\,dr |u_l(r)|^2= \int_0^{r_{c}}\,dr |u_{out}(r)|^2 +
\int_{r_{c}}^{r_{max}}\,dr |u_{in}(r)|^2=1
\end{equation}

Using the equations $(24)$

\begin{equation}
\int_0^{r_{max}}\,dr |u_l(r)|^2= (B f_{c})^2 \int_0^{r_{c}}\,dr |\Phi (r)|^2 +
 B^2\int_{r_{c}}^{r_{max}}\,dr |I(r)|^2
\end{equation}

 the normalization condition then reads

\begin{equation}
B^2 \biggl[ (f_{c})^2 \int_0^{r_{c}}\,dr |\Phi (r)|^2 +
\int_{r_{c}}^{r_{max}}\,dr |I(r)|^2\biggr]=1
\end{equation}

Denoting the result of the above integrals within brackets as $N$,
 we can write $B=\frac{1}{\sqrt{N}}$, thereby deriving the normalized eigenfunctions

\begin{equation}
 u_{out}(r)= \frac{1}{\sqrt{N}}\  f_{c}\  \Phi (r)
\end{equation}

\begin{equation}
 u_{in}(r)= \frac{1}{\sqrt{N}}\  I (r)
\end{equation}

\section{Integration and expectation values}

When performing the
 integration with the \emph{QQ-onia} package, we  use
 the following procedure: If we name

 \[T_N\ \equiv\ \int_{r_0}^{r_N}\,dr f(r)\   ;\  \\
 f_n\equiv f(r_n)\
 \]
  from the Euler-McLaurin summation formula
 \cite{abramov} with a given step $h$

\[
T_N = h \biggl[\frac{f_0}{2}+f_1+f_2+...+f_{N-1}+\frac{f_N}{2}\biggr]-
\frac{B_2 h^2}{2!}(f'_N-f'_0)-
\frac{B_{2k}h^{2k}}{(2k)!}(f^{(2k-1)}_N-f^{(2k-1)}_0)
\]
where $B_{2k}$ are Bernouilli numbers. The first term of the r.h.s.
 in the
 above equality corresponds to
 the \emph{extended trapezoidal rule}.\\

If we set a number of steps to a \emph{multiple of} 4, and apply the above
formula for steps h, 2h, and 4h we obtain

\[
T_h\approx h \biggl[\frac{f_0}{2}+ f_1+ f_2+ f_3
+...+\frac{f_N}{2}\biggr]- \frac{h^2}{12}(f'_N-f'_0)-
\frac{h^{4}}{4!30}(f^{'''}_N-f^{'''}_0)+ \textit{0}(h^6)
\]

\[
T_{2h}\approx 2h
\biggl[\frac{f_0}{2}+f_2+f_4+...+f_{N-2}+\frac{f_N}{2}\biggr]-
\frac{(2h)^2}{12}(f'_N - f'_0)-
\frac{(2h)^{4}}{4!30}(f^{'''}_N-f^{'''}_0)+ \textit{0}(h^6)
\]

\[
T_{4h}\approx 4h
\biggl[\frac{f_0}{2}+f_4+f_8+...+f_{N-4}+\frac{f_N}{2}\biggr]-
\frac{(4h)^2}{12}(f'_N - f'_0)-
\frac{(4h)^{4}}{4!30}(f^{'''}_N-f^{'''}_0)+ \textit{0}(h^6)
\]

where $T_N=T_{h}=T_{2h}=T_{4h}$. If we solve this system  to eliminate derivatives up to $\textit{0}(h^6)$
we arrive at the final formula of

\begin{equation}
\int_{r_0}^{r_N}\,dr\  f(r) \approx \frac{64 T_{h} - 20 T_{2h} + T_{4h}}{45}
+ \textit{0}(h^6)
\end{equation}

that is used in our calculations.\\

Several expectation values are needed to be computed in our method. 
Once the eigenfunctions have been normalized, one can calculate the
expectation value of a given operator $\textit{O}$ according to
the definition

\begin{equation}
\langle\  \textit{O}\  \rangle= \int_0^{r_{max}}\,dr\ u^{*}_l(r)\
\textit{O} \ u_l(r)
\end{equation}

 by using the above $\textit{0}(h^6)$ integration. $\textit{O}$ can be  $r^2$
 (if we want to obtain the mean square
 radius of the state $\sqrt{\langle r^2 \rangle}$), the potential or
 the derivative of the potential, etc.

\subsection{Square of the radial wave function at the origin}\

 We need to distinguish between the calculation of 
the square radial wave function at
 the origin \emph{(WFO)} for $l=0$ states, $|R_{n}(0)|^2$, and the
 calculation of the squared derivatives of the radial wave function at
 the origin for $l\neq 0$ states, $|R^{(l)}_{n}(0)|^2$, with
 $l=1,2,3$ respectively corresponding to the $P,D$ and $F$ states.

\subsubsection{$l=0$ states}\

From the well known calculation \cite{schob} derived from the
 Schr\"{o}dinger equation we obtain

\begin{equation}
|\Psi_{nlm_{l}}(0)|^2 = \frac{\mu}{2\pi}\  \langle\  V'(r)\  \rangle
\end{equation}

 where $\langle V'(r)\rangle$ is the expectation value of the
 derivative of the potential, and  $\mu$ stands for the reduced mass.
 If we are dealing with $b\overline{b}$ or $c\overline{c}$ systems:
 $\mu=m_Q/2$ where $m_Q$ is the heavy quark mass.\\

For $l=0$ states the wave function is: $\Psi_{n00}(r)=\frac{1}{\sqrt{4\pi}}\  R_{n0}(r)$, then

\begin{equation}
|R_{n}(0)|^2 = m_Q\  \langle\  V'(r)\  \rangle
\end{equation}

 Therefore, since the corresponding  eigenfunctions have been
already obtained, the evaluation of the radial \emph{WFO} reduces to a
 calculation that gives the expectation value of the derivative 
of the potential $\langle V'(r)\rangle$.\\

Another method that can be used to evaluate the \emph{WFO} is to
 extrapolate directly from the normalized eigenfunctions, taking
 values of the square $|R_{n}(r)|^2= [\ |u(r)|^2/r^2\ ]$ from the
 region near to $r=0$ using an appropriate interpolating function. Even so,
 the previously stated technique yields a  more accurate result, if
 the $\langle V'(r) \rangle$ calculation is reliable. Nevertheless
 both procedures can be employed together as a check.

\subsubsection{$l \neq 0$ states}\

Here we consider the \emph{SRE}, eq.(6), containing the centrifugal
 barrier term. If we rename the potential:

\[
W(r)\equiv V(r)+ \frac {\hbar^2 l(l+1)}{2\mu r^2}
\]
the kernel of the \emph{SRE} (see eq.(10)) 
now becomes: $k(r)\equiv \frac{2\mu}{\hbar^2}\ [E-W(r)]$.\\

\begin{description}
\item[\emph{l=1\ CASE}:]

 The main goal here is to evaluate the square
 first derivative at $r=0$, $|R'_{n}(0)|^2$.
 Since we have a tool for directly calculating the derivatives $u'(r)$,
 according to  eq. $(15)$, once we have the normalized eigenfunctions
  $u(r)$, then using

\begin{equation}
R'_{nl}(r)=\biggr[\frac{u(r)}{r}\biggl]'= \frac{u'(r)\ -\
R(r)}{r}
\end{equation}

we can extrapolate the square of the derivative, taking those values
 from the region near to $r=0$ with an appropriate
 interpolation function.

\item[\emph{l=2\ CASE}:]

 We will obtain the second derivative at $r=0$, $|R^{''}_{n}(0)|^2$  but \emph{only} in
 terms  of its radial  wave function and first derivative.
  To do this we proceed as follows: from \emph{SRE}

\begin{equation}
u^{''}(r)= -\ k(r)\ u(r)
\end{equation}

If we calculate the second derivative using the above identity

\begin{equation}
R^{''}_{nl}(r)=\biggr[\frac{u(r)}{r}\biggl]^{''}=\ -\ k(r)\ R(r)\ -
\frac{2\ R'(r)}{r}
\end{equation}

we can again extrapolate the $|R^{''}_{n}(0)|^2$ function.
 The main advantage of this procedure
is that there is no need for any additional derivatives, and
since $u(r)$ and $u'(r)$ have already been calculated
 at the suitable order, the result holds with the foreseen precision.

\item[\emph{l=3\ CASE}:]

In the same way, using eq. $(36)$, we can obtain the third derivative to extrapolate

\begin{equation}
R^{'''}_{nl}(r)=\biggr[\frac{u(r)}{r}\biggl]^{'''}= \biggr(\
\frac{2\ k(r)}{r}\ -\ k'(r)\ \biggl)\ R(r)\ +\ \biggr(\
\frac{6}{r^2}\ -\ k(r)\ \biggl)\ R'(r)
\end{equation}

where $k'(r)$ stands for the derivative of the kernel, $k'(r)=\ - \frac{2\mu}{\hbar^2}\ W'(r)$.
\end{description}

\subsection{Calculating the heavy-quark velocity }

To obtain the $\langle\  v^2\  \rangle$ value  we perform a double calculation as a check: the first
 is from the Hamiltonian definition and the second  uses the virial theorem.\\

\begin{description}
\item[i)] If $\vec{r}=\vec{r_1}-\vec{r_2}$ is the relative radial
 coordinate between the quarks 1 and 2, with velocities
 $|\vec{v_1}|=\ |\vec{v_2}| \equiv |\vec{v_q}| $, its relative velocity
 $\vec{v}$ at the center of mass frame is\newline
 $ \vec{v} = 2 \vec{v_1} = - 2 \vec{v_2}$. Then, we can obtain the quark velocity using the Hamiltonian,\newline
  $E=\langle T \rangle + \langle V(r)\rangle$, (where $T$ represents the relative kinetic
  energy $T = \frac{1}{2} \mu     (\vec{v})^2$)

\begin{equation}
\langle\ (\vec{v_q})^2\ \rangle = \frac{1}{2 \mu} \biggl[E- \langle\
V(r)\ \rangle \biggr]
\end{equation}

\item[ii)] Taking relative spherical coordinates, the virial theorem
implies

\[
\langle\  T\  \rangle = \frac{1}{2}\  \langle\  r\ V^{'}(r)\ \rangle
\]
Then, we derive the quark velocity from the expectation value
 of the product $r  V^{'}(r)$ according to:

\begin{equation}
\langle\ (\vec{v_q})^2\ \rangle = \frac{1}{4 \mu}\ \langle\  r\
V^{'}(r)\ \rangle
\end{equation}

\end{description}
in both calculations for $b\overline{b}$ or $c\overline{c}$ systems 
where $2 \mu = m_Q$.

\newpage

\begin{table}[htb]
\caption{$A_J$ and $B_J$  coefficients.}
\label{table:11}
\newcommand{\m}{\hphantom{$-$}}
\newcommand{\cc}[1]{\multicolumn{1}{c}{#1}}
\renewcommand{\tabcolsep}{2pc} 
\renewcommand{\arraystretch}{1.2} 
\begin{tabular}{@{}lll}
\hline
 $j$ value & $A_J$ & $B_J$ \\
\hline
$^3P_0$ & $-(16/3)$ & $+1$    \\
$^3P_1$ & $-(4/3)$  & $+(1/2)$\\
$^3P_2$ & $+(28/15)$& $-(1/2)$\\     
\hline
\end{tabular}\\[2pt]
\end{table}

\section{Bottomonia mass level splittings}\

In this section we address different splittings of the
mass levels of bottomonia in accordance with the expressions
shown in section \textbf{2.1}:

\begin{description}

\item[$n^3S_1\ -\ n^1S_0$ splitting:] In this case the only term that 
gives a non vanishing contribution is the 
spin-spin term. Since from \emph{QQ-onia} package we can previously 
calculate the $nS$ state WFO, instead of eq.(4) we choose to use 
its final and  well-known form  \cite{quigg2}

\begin{equation}
\Delta \biggl[\ M(^3S_1)-\ M(^1S_0)\ \biggr]\ =\  \alpha_s(\mu^2)\  \frac{8}{9m^2}\  |R_{n0}(0)|^2
\end{equation}

where we will evolve 
the $\alpha_s(Q^2)$  to the appropriate scale \cite{kiselev2} 

\[ Q^2\ =\ \langle \textbf{P} ^2 \rangle\  =\  2\ \langle\ T\ \rangle\  \mu\ =\  m_b\ \langle v^2 \rangle  \]

where we know the quark-velocity $\langle  v^2 \rangle $ 
for each $nS$ state from \emph{QQ-onia}.

\item[$n^3P_J$ and $n^1P_1$ splitting:] We have  to take into account 
altogether the spin-orbit and tensor terms, i.e. eqs. (3) and (5). 
Using a funnel potential, the explicit expressions to be computed read

\begin{equation}
\Delta \biggl[\ M(^3P_J)-\ M(centroid)\ \biggr]=\ \frac{1}{m^2}\ \biggl[A_J\  \alpha_s\ \hbar^3\  \langle\ \frac{1}{r^3}\ \rangle\ +\ B_J\  \sigma\ \hbar^2\ \langle\ \frac{1}{r}\ \rangle\ \biggr]
\end{equation}

where $A_J$ and $B_J$ are the corresponding coefficients for each case, 
as can be seen in Table 3; $\alpha_s$ denotes again the evolved value  
up to the quarkonium scale. The expectation values are 
calculated using the wave function $\forall r$ corresponding to the $nP$ centroid  from a previous calculation, as we will see later. Once  the $^3P_J$ masses are calculated, the   $^1P_1$ value will be obtained 
according to \cite{quigg2}, as 

\begin{equation}
M(^1P_1)\ =\ \biggl[\ 5\ M(^3P_2)\ +\ 3\ M(^3P_1)\ +\ M(^3P_0)\ \biggr]
\end{equation} 

\end{description}

\section{Electric Dipole Transitions}

\emph{QQ-onia} package allows us to calculate the wave functions of different states.   
Therefore one can calculate, e.g., the $E1$ transition 
rate $initial\ (i)\rightarrow final\ (f)\ +\ \gamma$ through 
the well-known expression \cite{quigg2}  

\begin{equation}
\Gamma_{E1}(i\rightarrow f\ +\ \gamma)\ =\ \frac{4\ \alpha\ \langle  e_Q \rangle ^2}{27}\ k^3\ (2J_f+1)\ S_{if}\ |\langle\ f\ |\ r\ |\ i\  \rangle|^2
\end{equation}  
where $\alpha=1/137$, and $\langle  e_Q \rangle $ is the mean
charge [ $(-1/3)$ in the bottomonia case ]. In the \emph{QQ-onia} file-example we will focus on $ ^3S_1\rightarrow ^3P_J$ radiative transitions with $S_{if}=1$. 
The photon energy, $k$ is directly calculated from energy-momentum conservation law. If $m_f$ and $m_i$ stands for the experimental \cite{pdg2} masses of 
the final resonance ($nP_J$) and the initial one ($nS$), respectively,  
one has

\begin{equation}
k\ =\ \frac{m_i^2\ -\ m_f^2}{2\ m_i}
\end{equation}

$\langle\ f\ |\ r\ |\ i\  \rangle$ (in GeV$^{-1}$) is the matrix element connecting final and initial state; we will evaluate it according to

\begin{equation}
\langle\ f\ |\ r\ |\ i\  \rangle\ =\  \int_0^{r_{max}}\,dr\ u^{*}_{f}(r)\ r \ u_{i}(r)
\end{equation}

extracting the reduced wave functions of the $nP$ and $nS$ states from a previous calculation with \emph{QQ-onia}.

\section{The \emph{QQ-onia} package}

The \emph{QQ-onia package} is written with \emph{PAW} software
 (\emph{Physics Analysis Workstation}), which can be obtained for free
 from the \textit{CERN} web site \cite{CERNpaw} for several operative
 systems. This software contains a FORTRAN interface, called SIGMA. The
 \emph{QQ-onia} package provides a version of this software named
 \emph{PAW-NT} for \emph{WINDOWS}. The package also runs with \emph{UNIX-LINUX},
  pasting  the \emph{bbnia-nl.kum} files directly into any \emph{UNIX} text editor.\\

The package contains the files prepared to work with each of the previously mentioned cases.
 To illustrate how the machinery runs, and as a potential reference,
 we choose a standard static potential for heavy quarkonium:
 the Coulomb plus linear potential \cite{eitchten2},
 and also as a further reference, we take a known set of parameters from
\cite{eitchten2}, \cite{quigg} with which to perform spectroscopy (for a comparison
of the Cornell model with other approaches of the heavy quarkonium potential in the
static limit, see \cite{kiselev}).\\

The files solve the \emph{SRE} for: $l=0$ states $\Upsilon(nS)
 (n=1,2,3,4)$, and their file names are \emph{bbnia-ns.kum}; the $l=1,\
 \chi(nP) (n=1,2)$  states, (\emph{bbnia-np.kum}); and the $l=2,\
 \Upsilon(nD)$ ones, (\emph{bbnia-1d.kum}). The package also contains
 a file called \emph{bbnia-4f.kum} which is prepared for working with theoretical
 $l=3\ (nF)$ states. The differences between these files is related to the
 centrifugal barrier and the calculation of either square of the radial wave functions
 or square of the derivatives at the origin. Since there are parts that
 are common to both we will begin by explaining them.\\

 It must be pointed out here that in some parts of the software it is necessary to type
 in some analytic expressions (such as the derivative of the
 potential and the product $rV'(r)$). Although this procedure  can be
 shortened, this  is a good way to ensure control
 over the different variables, useful to analyze partial results.

\section{Getting ready to calculate energy levels: general settings}\

First we open the \underline{file section \texttt{SETTINGS}} in
which we can read (to insert) the characteristic items:\

\begin{description}
\item[\texttt{v/cr igma(1) r 0.9255}] creates and defines the string tension (in $GeV/fm$).
\item[\texttt{v/cr mq(1) r 5.18}] creates and defines the b-quark mass.
\item[\texttt{v/cr alf(1) r 0.39}] creates and defines the strong coupling
  constant.
\item[\texttt{v/cr hb(1) r 0.19732858}] is the $\hbar c$ constant (in $GeV fm$ units).
\item[\texttt{ni}] sets the number of total steps; \texttt{h=0.001} ($fm$) is the step.
\item[\texttt{x0}] sets the minimum $r$ value, $r_{min}$: the \texttt{h} value is the default.
\item[\texttt{xm=[x0]+([ni]-1)*[h]}] sets the maximum $r$ value, $r_{max}$
\item[\texttt{xc}]sets the cut $r$ value $r_{cut}$ according to $B_{est}= V(r_{cut})$, where
 we use $B_{est}$ as an estimated binding energy  $B_{est}= M_{exp}-
 2 m_b$, and  $M_{exp}$ stands for the experimental mass of the
 resonance. Sometimes, if needed, we can also set $r_{cut}$ as the point
 at which the inwards integration has its first maximum.

\item[\texttt{v/cr ele(1) r ''value''}] (with \texttt{ele}$=1,2,3$) \emph{only} appears in
 $l\neq 0$ files, where it creates and defines the eigenvalue of the angular momentum that
 needs to be inserted into the centrifugal barrier term.
\end{description}\

\underline{File section  \texttt{SIGMA APPLICATION BLCK1}} calculates the $r$ range and defines the potential:
\begin{description}
\item[\texttt{x=array([ni],x0\#xm)}] it calls \texttt{sigma application (appl sigma)}
  to establish the $r$ range. The \underline{numerical} values of \texttt{ni,x0,xm}, must be inserted here.
\item[\texttt{POTENTIAL DEFINITION}] \underline{for $l = 0$ files} The $V(r)$
value, (\texttt{vo} variable) is calculated for the full $r$-range. In our case \texttt{vo=(igma*x)-(kfac1*(x**(-1)))}.
\item[\texttt{POTENTIAL DEFINITION}]  \underline{for $l\neq 0$ files};\newline
 \texttt{vo=(igma*x)-(kfac1*(x**(-1)))+(bfac2*(x**(-2)))}, which contains the centrifugal barrier term, sets the $W(r)$ value.
\item[\texttt{ENERGY  SETTINGS}] The programme calls the \texttt{sigma application (appl sigma)}
  to establish the energy range (using the variable \texttt{e}); here the \underline{numerical} values
  of\newline
  \texttt{[ne],E-min,E-max}  must be inserted as follows:
 \newline
 \texttt{e=array(}number of energy steps, minimum energy value\texttt{\#} maximum energy value\texttt{)}.
\item[\texttt{ENERGY  BLCK2}] to insert the number of energy steps (\texttt{ne}). The programme will calculate
  the corresponding energy values, \texttt{e}, to cover the full energy range.

\item[\texttt{BOUNDARY CONDITIONS}] this section of the programme runs automatically,
the default values are those previously
discussed in section \textbf{2.2} $u(r\rightarrow 0) \sim r^{l+1}$ and
$u(r\rightarrow \infty) \sim \exp \biggl[ - \frac{\sqrt{2\mu
|B_{est}|}}{\hbar}r \biggr]$.\\

those conditions are inserted into the two first values of $u_{out}(r)$ (\texttt{uo} variable) and
$u_{in}(r)$ (\texttt{ui} variable), respectively

\item[In summary],we set  \texttt{h, [ni], x0, xm, xc, [ne], E-min, E-max},
the potential, and the corresponding constants.

\end{description}

\subsection{Starting and determination of the bound state energy level}\

Once we have inserted the previous settings, we write \texttt{exec bbnia-nl.kum}\ in the \emph{PAW}
 interface. The programme then automatically searches for bound states within the $[E_{min},E_{max}]$ range.
 If there is any change of sign in the $G(E)$ function the programme stops, showing the new energy
  range $[E'_{min},E'_{max}]$ in which the bound state can be found, and it asks for the next
   instruction \newline
    (\texttt{Type <CR> to continue or Q to quit}). We then proceed in an iterative way:
    to improve the precision of the energy level, we skip the programme (by typing in \texttt{q})
    and on opening the file, we type in the new energy range in the \texttt{ENERGY SETTINGS} file section.
    We then restart the programme and repeat this process until the energy level has the desired
    tolerance.\\

  Once the programme stops showing a suitable energy range $[E^{last}_{min},E^{last}_{max}]$,
  we push the \texttt{enter} key. The programme will then ask for the \emph{final} (\texttt{ee(1)})
  energy eigenvalue, then we type in this  value (which could be the mean value
  of $[E^{last}_{min},E^{last}_{max}]$), and it will calculates automatically the
  eigenfunctions $\forall r$, showing at the screen the normalization proof.\\

As a final result we get:
\begin{description}
\item[1] The normalized reduced radial wave function, $u(r)$ throughout the full range, stored as a
\texttt{[ni]}-dimensional vector (\texttt{y1})

\item[2] The $u^*(r)\ u(r)$ product (\emph{in $fm^{-1}$ units})which is also stored as
a \texttt{[ni]}-dimensional vector (\texttt{fc}). These values  can
be either displayed in a  graph, by typing in \texttt{v/dr fc}, or obtained
 numerically (\texttt{v/wr fc}).
\end{description}

\subsection{Mean square radius}\

\underline{File section \texttt{**BEGINS SQR(<r2>) CALCULATION**}}.
 The software automatically calculates the mean square radius of the
 state ($\sqrt{\langle r^2 \rangle}$) according to its expectation
 value. The \texttt{[ni]}-dimensional variable (\texttt{fi}) to
 integrate is \texttt{fi=fc*xx}, where \texttt{xx} is the $r^2$
 variable. As this is the final result, this value is stored  at
 the \texttt{rad} variable (in $fm$ units), and displayed.

\subsection{Computing the heavy-quark velocity using the virial theorem}\

\underline{File section  \texttt{**1) <v2>/c2 calculation USING VIRIAL THEOREM**}}

\begin{description}
\item[1] The first step is to calculate the expectation value of the product of
$X\cdot \nabla V$, $(\ \langle\  r V'(r)\  \rangle\ )$; to do this, \emph{BEFORE running the programme},
we must \emph{TYPE} in its corresponding expression at the variable \texttt{xdpot}. Using our example
 we must insert:\newline
\texttt{xdpot=(igma*x)+(kfac1*(x**(-1)))}

\item[2] The programme  integrates the variable \texttt{fi=fc*xdpot} and performs
the velocity calculation. The result is displayed and stored at the
variable \texttt{v2}.

\end{description}

\subsection{Computing the heavy-quark velocity using the Hamiltonian}\

\underline{File section \texttt{**2) <v2>/c2 calculation USING <H>=<T>+<V(r)>**}}.

\begin{description}
\item[1] The programme first calculates the expectation value of the potential $\langle V(r) \rangle$.
The variable to integrate is \texttt{ffi}, where \texttt{ffi=vo*fc}. The
 result is stored at the variable \texttt{vbar}. If
 desired, it can be read by dropping the comment variable
\texttt{*} at line:
\newline
\texttt{*v/pr vbar}.

\item[2] It calculates the velocity value using the variables: \texttt{vbar},
 the energy eigenvalue, \texttt{ee}, and the inverse of the
 quark mass, \texttt{inq}, and displays the final result.
\end{description}

\section{Computing the radial wave function squared 
and derivatives at the origin}

\begin{description}

\item[Square of the radial \emph{WFO}] (For $l=0$ states:
 \texttt{bbnia-ns.kum}\ files).\newline
 \underline{File section  \texttt{**BEGINS WAVE FUNCTION AT THE ORIGIN CALCULATION**}}.
 First  the expectation value of the derivative of the potential
  $\langle V'(r) \rangle$ is calculated; to do this, \emph{BEFORE the programme runs}, we must type in
   its corresponding expression at the variable \texttt{dpot}.
   Using our example, we  must insert:
 \newline
\texttt{dpot=(igma)+(kfac1*(x**(-2)))} \newline
 The variable to integrate is \texttt{fi}, where \texttt{fi=dpot*fc}. Later, the final calculation
 is performed and  the result is displayed (in GeV$^3$ units) and stored
  at the \texttt{wfo} variable.

\item[Square of the first derivative of the radial \emph{WFO}]
 (For $l=1$ states:\newline \texttt{bbnia-np.kum}\ files). \newline
 \underline{File section \texttt{WAVE FUNCTION DERIVATIVE CALCULATION}}.
  General  settings are equal than before by changing $V(r)\rightarrow
 W(r)$, then  the centrifugal barrier term appears\newline
 \texttt{(bfac2*(x**(-2)))}.
 As a result we obtain, according to eq.(35), the  squared  first derivative of the radial wave function $|R'_{n}(r)|^2$
 \emph{for all r value} (in GeV$^5$ units). The result is stored at the
 \texttt{[ni]}-dimensional variable  \texttt{derc}. Then, to obtain its value at the origin,
 we must export the (\texttt{derc}) values in order to to extrapolate with an appropriate tool (such as
  the \emph{PAW} inner routine \texttt{vector/fit}). By typing in \texttt{v/dr derc}, we can
 plot these values to select the range before exporting. To do this numerically
 type in \texttt{v/wr derc}.

\item[Square of the  second derivative of the radial \emph{WFO}]
(For $l=2$ states:\newline \texttt{bbnia-nd.kum}\ files). \newline
 \underline{File Section \texttt{WAVE FUNCTION DERIVATIVE(2) CALCULATION}}
 We obtain, by using the eq.(37), the second  derivative of the radial wave function
 (in GeV$^7$ units) $|R''_{n}(r)|^2$ values
  \emph{for all r}, these values are stored at the \texttt{[ni]}-dimensional
  variable \texttt{sed2}, and ready to be  extrapolated.

\item[Square of the third derivative of the radial \emph{WFO}] (For  theoretical $l=3$ states:\newline
\texttt{bbnia-nf.kum} files) \newline \underline{File section
\texttt{WAVE FUNCTION DERIVATIVE(3) CALCULATION}}. Before starting
the programme, we must \emph{TYPE} in
 the corresponding derivative $W'(r)$ expression into the variable \texttt{kerd0}.
 Then the programme will calculate
 the derivative of the kernel, $k'(r)$. As a result, according to eq.(38), we
 have the   \texttt{[ni]}-dimensional variable \texttt{thrd2}, which
 stores the full $r$-range of the third derivative
 of the radial wave function $|R'''_{n}(r)|^2$ (in GeV$^9$ units).
\end{description}

\subsection{Wave function plot} When a \texttt{bbnia-nl} file stops, a plot of
 the dimensionless reduced wave function can be obtained (from the same \emph{PAW} screen),
  by running the routine \texttt{graph-nl.kumac}. We obtain  the plot
   of the \texttt{u2} variable as output, which is the $|u(r)|^2$ value
  multiplied by the B\"ohr radius of the resonance, $a_0=\ \hbar/ (C_F\ \alpha_s\ \mu)$.
  
\section{\emph{QQ-onia} package spin-dependent and $E1$}

This part of the software is organized within \emph{QQ-onia} package as follows: The file \emph{SSplit-nS.kumac} allows to calculate the $n^3S_1\ -\ n^1S_0$ splitting for each $n$ level. The files \emph{Split-nP.kumac} ($n=1,2$) calculate as example the $n^3P_J$ and $n^1P_1$ splitting. There is also a file  \emph{E1-2S1P.kumac}, which analyzes  the $E1$ $2S\rightarrow1P$ transitions, which has two subroutines named \emph{ini-2s.kum} and \emph{final-1P.kum} devoted to generate, respectively, the initial and final states for the matrix element calculation.

\begin{description}

\item[SSplit-nS.kumac] This file has mainly three blocks: the first one ask for the values of the quark mass (the \texttt{mq} variable), the  value of the wave function at the origin (\texttt{wfo} variable), and the quark velocity of the $nS$ state (\texttt{v2q} variable), whose values can be previously found with  the  \emph{bbnia-ns.kum} files. Once these values are entered the programme runs automatically; it calculates the quarkonium scale and the $\alpha_s$ value (\texttt{alf} variable) through the \texttt{alpha-s Evolution Block}. Later using the eq.(41)\newline (the \texttt{Delta-ss=[M(n3S1)-M(n1S0)]calculation block}) it displays the result of the energy splitting (in $GeV$) 
 through the variable \texttt{deltss}.

\item[Split-nP.kumac] The first part of the file generates the centroid wave function,(the \texttt{GENERATION OF THE CENTROID WAVE FUNCTION-BLOCK}. This implies to run first the corresponding \emph{bbnia-np.kum} in order to know the energy eigenvalue, the $r_{max}$ and $r_c$ values to be inserted here [We \emph{insert} the energy eigenvalue (\texttt{ee} variable) when the programme ask for it (\texttt{EIGENVALUE?(GeV)})]. This block can be performed in other cases  by pasting from the \emph{bbnia} file the part corresponding to the block labeled \texttt{ONCE ENERGY LEVEL FOR BOUND STATE IS DETERMINED***************\newline WAVEFUNCTION CALCULATION}. At  the end of this part we have stored the normalized wave function within the \texttt{y1} variable. Later the programme will run automatically, it calculates the $<r^{-3}>$ and $<r^{-1}>$  expectation values, evolves the $\alpha_s$, and applies the final expression eq.(42). Finally it displays the energy differences and the  final $^3P_J$ masses [Also the $^1P_1$ mass using eq.(43)] by means of 
\[
M( ^3P_J )= \Delta E_J + M{centroid}
\]
 where $\Delta E_J$ is the calculated energy difference of each $J$ and  $M{centroid}$ is the centroid mass \emph{previously} calculated from the \emph{bbnia-np.kum} file.

\item[E1-2S1P.kumac] This file, as a complementary tool, is an example of how to handle \emph{QQ-onia} with the E1 transition rates between $2S$ and the $^3P_J$ states. First, the programme calls the \emph{ini-2s.kum} subroutine in order to create the initial $2S$ state, which is performed by inserting the results found using the \emph{bbnia-2S.kum} file: the energy eigenvalue (at the \texttt{Enter the energy eigenvalue} line), the $r_{max}$ and $r_c$ values; then it generates its corresponding wave function. Later, the software calls the \emph{final-1p.kum} subroutine and, in the same way, it creates the wave function corresponding to the final state. It must be pointed out that, when we want to modify the \emph{common} parameters used generating the  final and initial states, we perform it  within the  \emph{E1-2S1P.kumac} file. The programme then runs automatically: it calculates, using eq.(46), the matrix element $<f|r|i>$ (stored at the \texttt{me} variable in GeV$^{-1}$), the photon energy for each final $J$ (the \texttt{kph0,1,2} variables, in $GeV$) according to eq.(45), later it computes the final values, in $keV$, by means of the eq.(44), the result is displayed for each $J=0,1,2$ value through their respective \texttt{gma0,1,2} variables. The remaining factors and settings are explained and displayed inside the \emph{E1-2S1P.kumac} file.
 
\end{description}

\section{Numerical accuracy}\

One important issue is determination of the numerical accuracy of the
results. Since no complete  bottomonium wave function has yet been
available, we can make our check  using an harmonic oscillator
potential, which has a known analytical
  solution, but it is  numerically sensitive (as previously mentioned, this potential
   exhibits numerical instabilities due to their exponentially growing solutions).
     It is therefore a good candidate for checking our software.\\

We test $V(r)= \frac{1}{2} \mu\ \omega^2\  r^2$ by inserting it in the
file block \texttt{SIGMA APPLICATION BLCK1}, and by inserting in the
\texttt{SETTINGS} file section $\mu =100\ \mathrm{MeV}/c^2$;\  
$\hbar c = 197.32858\  \mathrm{MeV}
fm$;\  $\omega = 2\  fm^{-1} c$. As boundary conditions \cite{gasio}
we use  $u(r\rightarrow 0) \sim r^{l+1}$ and for $u(r\rightarrow
\infty) \sim e^{-\beta r^2}$, where
$\beta =\frac{\mu\ \omega}{\hbar c}$.\\

The analytical energy levels are
 $E(n,l)= \biggr(2n+l+\frac{3}{2}\biggl)\hbar\omega $.\\

 The reduced wave functions expressed in terms of the generalized
  Laguerre polynomials \cite{gasio} are:\\

 \begin{equation}
 u_{nl}(r) =\ \biggl[\ \frac{2n!\alpha^3}{\Gamma(n+l+3/2)}\  \biggr]^{1/2}\  r\  (\alpha r)^l\  L^{l+1/2}_n
 (\alpha^2 r^2)\  e^{-\alpha^2 r^2/2}
 \end{equation}

We then analyze levels $1S\equiv(0,0); 2S\equiv(1,0)$ and $2D\equiv(1,2)$.
The results can be found in Table 4.

\begin{table}[htb]
\caption{Harmonic oscillator results.} \label{table:3}
\newcommand{\m}{\hphantom{$-$}}
\newcommand{\cc}[1]{\multicolumn{1}{c}{#1}}
\renewcommand{\tabcolsep}{2pc} 
\renewcommand{\arraystretch}{1.2} 
\begin{tabular}{@{}lll}
\hline
Energy level (MeV) & Analytic & Numeric  \\
\hline
$E(0,0)$        & 591.986& 591.999 \\
$E(1,0)$        & 1381.330& 1381.315 \\
$E(2,1)$        & 2170.614& 2170.628 \\
\hline
\end{tabular}\\[2pt]
\end{table}

\begin{description}

\item[(0,0)LEVEL]

Taking $r_{cut}$ as the turning point (using $E(0,0)= V(r_{cut})$), we find that
 $r_{cut} = 1.72\  fm$ with $r_{max}=5\ fm$ and $r_{min} = h$, with a step $h=0.01\  fm$, thus
 $[ni]=500$. We start searching for the associated energy level through the range $E(0,0)\in [400,800] \mathrm{MeV}$,
 with an energy step of $\Delta_E=1$ MeV, i.e. $[ne]=401$.

\item[(1,0)LEVEL]

We repeat the procedure, but now $r_{cut} = 2.6\  fm$ with $r_{max}=5.6\  fm$, thus
$[ni]=560$; starting with an energy range $E(1,0)\in [1100,1500] \mathrm{MeV}$,
 and an energy step of $\Delta_E=1$ MeV.

\item[(1,2)LEVEL] First the centrifugal barrier term $ \frac {\hbar^2 l(l+1)}{2\mu r^2}$
must be added to \texttt{SIGMA APPLICATION BLCK1}. We use $r_{cut} =
1.08\ fm$, $r_{max}= 6\ fm$, with an energy range  \newline
 $E(1,2)\in [1900,2300] \mathrm{MeV}$, $\Delta_E=1\  \mathrm{MeV}$.

\end{description}

\begin{figure}[htb]
\begin{center}
\includegraphics[width=15pc]{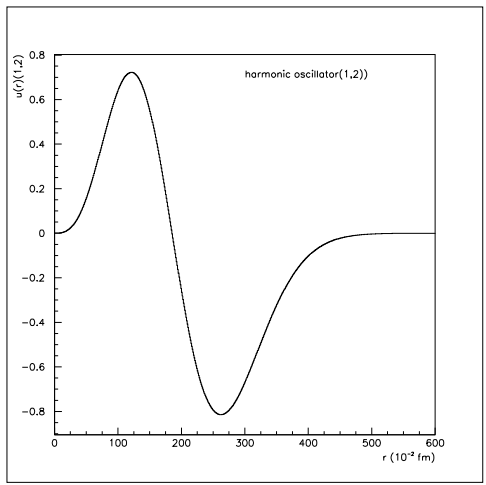}
\end{center}
\caption{Harmonic oscillator$(n=1,l=2)$ reduced wave function.}
\label{fig1}
\end{figure}

\begin{figure}[htb]
\begin{center}
\includegraphics[width=15pc]{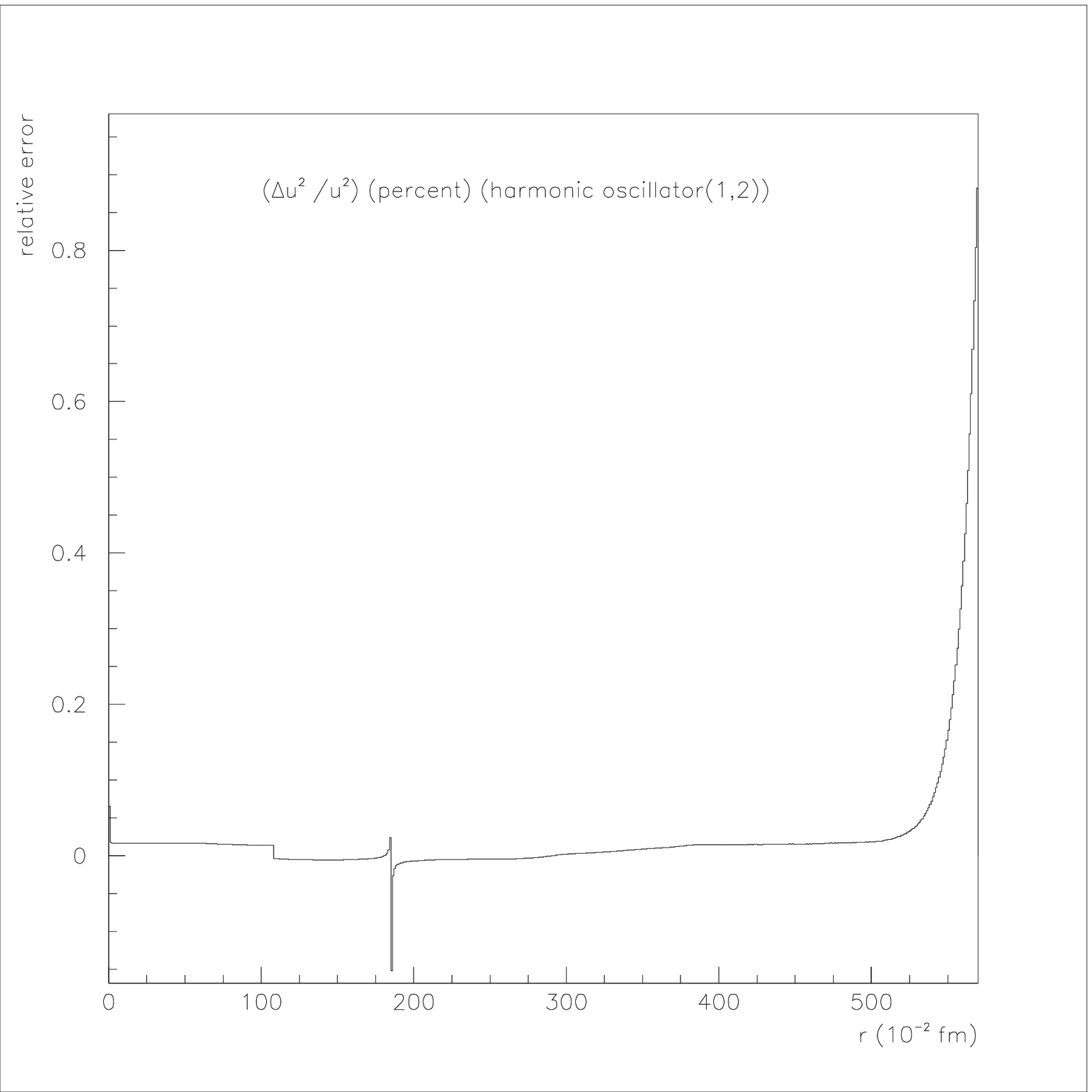}
\end{center}
\caption{Harmonic oscillator$(n=1,l=2)$ relative error.}
\label{fig2}
\end{figure}

Concerning the reduced radial wave functions $u(r)_{nl}$, in all
cases, we find a deviation of less than $1\%$ with respect to the
exact value from eq.$(47)$. Figure 1 shows the result for the
$u(r)_{12}$  case; Figure 2 shows  its corresponding $\Delta
|u(r)|^2_{12}/|u(r)|^2_{12}$ relative error (in $\%$).

\section{An example: Numerical Results}

To illustrate how to the \emph{QQ-onia} package runs, all attached
\texttt{bbnia-nl.kum}\  files can be opened; their
 corresponding settings are listed inside.
 Here, we will focus on the $\Upsilon(1S)$ case (\texttt{bbnia-1s.kum}\ file)
  in order to explain some relevant details.\\

We look for the lowest state of the $b\overline{b}$ family. First we
estimate its energy eigenvalue, $B^{1S}_{est}$, from the
experimental data \cite{pdg2}, then $B^{1S}_{est}= M^{1S}_{exp}\ -\ 2
m_b= -0.8997\ GeV$, (with $M^{1S}_{exp}= 9.4603\ GeV$ and $m_b=5.18\
GeV$). To ensure that we find the lowest state
 we set a wide energy range  below the $B^{1S}_{est}$, thus we start typing at \texttt{ENERGY SETTINGS} section
 a trial energy range $B^{1S}\in [-5.,0.]$ GeV 
with an energy step of $\Delta_E =0.5$ GeV, i.e. \texttt{[ne]}$=11$.\\

 Using $B^{1S}_{est}= V(r_{cut})$, we find $r_c\approx0.1\ fm$ (alternatively,
 instead of $B_{est}= V(r_{cut})$, we can take
  the $r_c$ point where the inwards integration has its first maximum). To ensure the asymptotic conditions we
 then take $r_{max}= 1.0\ fm$; taking $h=0.001\ fm$, the number of steps is \texttt{[ni]}$= 1000$.\\

We type in \texttt{exec bbnia-1s.kum} at \emph{PAW} screen, then the
programme runs until it stops displaying: \newline

\texttt{¦bound state around¦\newline
E(10) = -0.5\newline
E(11) = 0\newline
Type <CR> to continue or Q to quit}\\

This is the new energy range (in GeV) within which we can find the
eigenvalue of the bound state.
  We then stop the programme (\texttt{q}) and  insert these
new values at the \texttt{ENERGY SETTINGS} section and restart the
programme. We repeat the procedure, and after five iterations the
screen displays: \newline

\texttt{¦bound state around¦\newline
E(10) = -0.170209\newline
E(11) = -0.170208\newline
Type <CR> to continue or Q to quit}\\

As there is sufficient precision, when we press the \texttt{enter}
key, the screen shows:\\

\texttt{ ¦BOUND STATE ENERGY?¦\newline
EE(1)}\newline

 We then take  the mean value of the above quantities
$-0.1702085$ (GeV) as the eigenvalue, and type it in at the
screen.\newline

\texttt{EE(1) -0.1702085}\\

As the calculated eigenvalue $B^{1S}_{calc}$ does not match the
$B^{1S}_{est}$ value, we can establish  a scale factor ($F$)
\emph{to be applied to the whole spectrum}:\newline
\[
F =\ M^{1S}_{exp}\ -\ (2 m_b\ +\ B^{1S}_{calc})
\]

(Alternatively, we can redefine the quark mass according to
 $2m'_b \equiv 2m_b + F$). The masses  of the higher states $M^{nl}_{calc}$ can
therefore be obtained from

\[
M^{nl}_{calc}=\ 2 m_b\ +\ B^{nl}_{calc}\ +\ F
\]

where, $B^{nl}_{calc}$ stands for the calculated energy eigenvalue
obtained from the software for each higher state.\\

In Table 5 the results for the masses of several
resonances can be found.

\begin{table}[htb]
\caption{Coulomb plus linear potential: mass results (in GeV
units).} \label{table:4}
\newcommand{\m}{\hphantom{$-$}}
\newcommand{\cc}[1]{\multicolumn{1}{c}{#1}}
\renewcommand{\tabcolsep}{2pc} 
\renewcommand{\arraystretch}{1.2} 
\begin{tabular}{@{}llll}
\hline
 $b\bar{b}$ LEVEL & Experimental mass & Mass from \cite{eitchten2} & Mass from \emph{QQ-onia}\\
\hline
$\Upsilon(1S)$ & $9.4603$   & $ 9.4603^{a}$& $9.4603^{a}$\\
$\chi(1P)^{b}$ & $9.9001$   & $ 9.96 $     & $9.9584$ \\
$\Upsilon(2S)$ & $10.02326$ & $ 10.05$     & $10.02772$ \\
$\Upsilon(1D)$ & $10.1622$  & $ 10.20$     & $10.2080$ \\
$\chi(2P)^{b}$ & $10.2620$  & $ 10.31$     & $10.3125$ \\
$\Upsilon(3S)$ & $10.3553$  & $ 10.40 $    & $10.3971$ \\
$b\bar{b}(4F)^c$& $ ----$    & $ ---- $     & $10.3995$ \\
$\Upsilon(4S)$ & $10.5794$  & $ 10.67 $    & $10.6739$ \\
\hline
\end{tabular}\\[2pt]
$(a)$ used to set the ground level in both references\\
$(b)$ $\chi_{bJ}(nP)$ centroid \   $(c)$ Theoretical level
\end{table}

After  inserting the energy eigenvalue \texttt{EE(1)}, the programme
 will calculate the remaining  quantities automatically: it first shows the
 normalization check and then displays a plot of the \texttt{fc} value. \\

\texttt{ ¦normalization proof¦\newline
NNOR(1) = 1}\newline

It continues to show the value  of the square of the wave function at the
 origin
\newline \texttt{¦RADIAL WAVEFUNCTION AT THE ORIGIN (in
GeV$^3$)¦\newline WFO(1) = 14.0927}\newline

\begin{table}[htb]
\caption{Coulomb plus linear potential: $|R^{(l)}_{n}(0)|^2$ values
in (GeV)$^{(3 + 2l)}$ units.} \label{table:5}
\newcommand{\m}{\hphantom{$-$}}
\newcommand{\cc}[1]{\multicolumn{1}{c}{#1}}
\renewcommand{\tabcolsep}{2pc} 
\renewcommand{\arraystretch}{1.2} 
\begin{tabular}{@{}lll}
\hline
 $b\bar{b}$ LEVEL & $|R^{(l)}_{n}(0)|^2$  from \emph{QQ-onia} &  $|R^{(l)}_{n}(0)|^2$  from \cite{quigg}\\
\hline
$\Upsilon(1S)$        & $14.09$& $14.05$ \\
$\chi(1P)$            & $2.062$& $2.067$ \\
$\Upsilon(2S)$        & $5.947$& $5.668$ \\
$\Upsilon(1D)$        & $0.835$& $0.860$ \\
$\chi(2P)$            & $2.440$& $2.438$ \\
$\Upsilon(3S)$        & $4.276$& $4.271$ \\
$b\bar{b}(4F)^c$       & $0.551$& $0.563$ \\
$\Upsilon(4S)$        & $3.675$& $3.663$ \\
\hline
\end{tabular}\\[2pt]
\end{table}

 If we run a $l\neq0$  file, we do not see this screen but \emph{when the programme finishes}
 we can export the corresponding  $|R^{(l)}_{n}(r)|^2 (\forall r)$
 and extrapolate.\\

Table 6 summarizes the results for the square of the \emph{WFO} (or its
 derivatives).

 After the \emph{WFO} value, the calculation  for the mean square radius calculation
   from the $1S$ file will show \newline
 \texttt{¦SQR(<$r^2$>) (in fm)¦\newline
 RAD(1) = 0.201043}\newline

\begin{table}[htb]
\caption{Coulomb plus linear potential: mean square radius (in
$fm$).} \label{table:6}
\newcommand{\m}{\hphantom{$-$}}
\newcommand{\cc}[1]{\multicolumn{1}{c}{#1}}
\renewcommand{\tabcolsep}{2pc} 
\renewcommand{\arraystretch}{1.2} 
\begin{tabular}{@{}lll}
\hline
 $b\bar{b}$ LEVEL & $\sqrt{\langle r^2 \rangle}$ from \emph{QQ-onia} &  $\sqrt{\langle r^2 \rangle}$ from \cite{eitchten2}\\
\hline
$\Upsilon(1S)$        & $0.20$ & $0.20$ \\
$\chi(1P)$            & $0.38$ & $0.39$ \\
$\Upsilon(2S)$        & $0.46$ & $0.48$ \\
$\Upsilon(1D)$        & $0.52$ & $0.53$ \\
$\chi(2P)$            & $0.63$ & $0.64$ \\
$\Upsilon(3S)$        & $0.71$ & $0.72$ \\
$b\bar{b}(4F)^c$       & $0.64$ & $----$ \\
$\Upsilon(4S)$        & $0.91$ & $0.92$ \\
\hline
\end{tabular}\\[2pt]
\end{table}

  The results for each resonance can be seen in Table 7.\\

 Finally, the $1S$ file displays the velocity result obtained applying the two
 methods\\

 \texttt{¦(<$v^2$>/$c^2$) [virial theorem] ¦\newline
 V2(1) = 0.0962335\newline
 ¦(<$v^2$>/$c^2$) [<H>=<T>+<V>] ¦\newline
 V2A(1) = 0.0962476}\\

and the corresponding results for each case can be seen in Table 8.\\

 Finally, as an illustrative example, 
one can observe in Figure 3 the dimensionless $\Upsilon(1S)$
 reduced wave function taken from the \emph{graph-nl.kumac} file.

\begin{figure}[htb]
\begin{center}
\includegraphics[width=15pc]{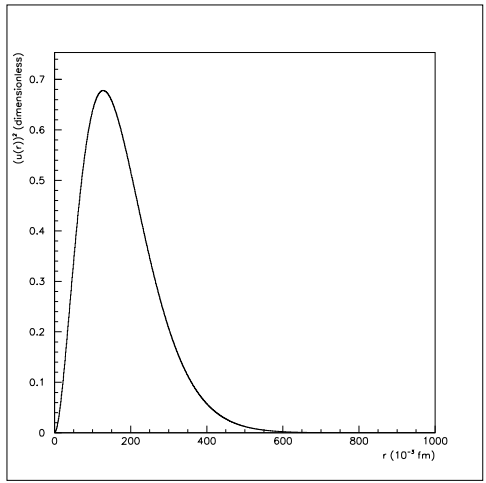}
\end{center}
\caption{$\Upsilon(1S)$ reduced wave function.} \label{fig3}
\end{figure}

\begin{table}[htb]
\caption{Coulomb plus liner potential: velocities $\langle\  v^2\  \rangle$.}
\label{table:7}
\newcommand{\m}{\hphantom{$-$}}
\newcommand{\cc}[1]{\multicolumn{1}{c}{#1}}
\renewcommand{\tabcolsep}{2pc} 
\renewcommand{\arraystretch}{1.2} 
\begin{tabular}{@{}lll}
\hline
 $b\bar{b}$ LEVEL & $\langle\  v^2\  \rangle$ from $\emph{QQ-onia}^a$ &  $\langle\  v^2\ \rangle$ from \cite{eitchten2}\\
\hline
$\Upsilon(1S)$        & $0.096$& $0.096$ \\
$\chi(1P)$            & $0.065$& $0.065$ \\
$\Upsilon(2S)$        & $0.078$& $0.076$ \\
$\Upsilon(1D)$        & $0.067$& $0.067$ \\
$\chi(2P)$            & $0.075$& $0.076$ \\
$\Upsilon(3S)$        & $0.085$& $0.085$ \\
$b\bar{b}(4F)^c$       & $0.073$& $-----$ \\
$\Upsilon(4S)$        & $0.096$& $0.097$ \\
\hline
\end{tabular}\\[2pt]
$a:$ Results doubly checked from Hamiltonian and Virial methods.
\end{table}


\begin{table*}[hbt]
\setlength{\tabcolsep}{0.4pc}
\caption{Values of the predicted and experimental masses (in GeV),
square of the WFO (or derivative) in GeV$^{3+2l}$, mean square radius (in
fm) and the typical quark velocity for the $\Upsilon(1S,2S,3S,4S)$,
$\chi_b(1P,2P)$ and $\Upsilon(1D)$ states for the Cornell-modified
potential.
\newline}

\label{table:8}

\begin{center}
\begin{tabular}{lccccc}
\hline Resonance & Mass & Exp. & $|R_{nl}^{l}(0)|^2$ & $\langle r^2
\rangle^{1/2}$ &
$\langle v^2 \rangle$  \\
\hline
$\Upsilon(1S)$ & 9.4603 & 9.4603  & $12.65$ & $0.23$ & $0.094$  \\
\hline
$\chi_b(1P)$ & 9.8929 & 9.9001  & $1.409$ & $0.40$ & $0.071$  \\
\hline
$\Upsilon(2S)$ & 10.0236 & 10.0233 &  $6.444$ & $0.51$ & $0.091$    \\
\hline
$\Upsilon(1D)$ & 10.1476 &  10.1622 & $0.562$ & 0.53 & $0.078$  \\
\hline
$\chi_b(2P)$ & 10.2729 & 10.2600  & $1.854$ & $0.63$ & $0.089$  \\
\hline
$\Upsilon(3S)$ & 10.3750 & 10.3552 & $5.404$ & $0.71$ & $0.103$  \\
\hline
$\Upsilon(4S)$ & 10.6477 &  10.5794 & $5.194$ & 0.88 & $0.120$  \\
\hline
\end{tabular}
\end{center}
\end{table*}

\section{More results using another potential}

As a further example, we present  results obtained using a Leading
Order potential for a heavy quarkonia system
\[
V^{{\mathrm{LO}}}=V^{(0)} + \frac{V^{(1)}}{m_b}
\]

which contains a $V^{(0)}$ static term (which could be a Coulomb
plus linear potential), and an additional \textit{O}$(1/m)$ piece  whose
contribution is  comparable to the static part. From Lattice
analysis \cite{Koma:2006si,Koma:2007jq}, we set the explicit form of
this potential as follows (the so called Cornell-modified potential,
see \cite{Domenech-VV1} and references therein for a complete
description):

\begin{equation}
V_{Cor-mod}(r)=-\frac{c}{r}-\frac{c'}{r^2}+\sigma r + \mu
\end{equation}

The values of  parameters $m_b,c,c',\sigma$ and $\mu$ were obtained
by applying a fitting procedure to the bottomonium spectrum (using
$\Upsilon(1S)$ and $ \Upsilon(2S)$ states), not from Lattice
estimates. From the fit we obtained  the following values for the
parameters of the potential:\newline $m_b=4.7$ GeV, ($\mu=0$)

\[ \sigma=0.217\ {\mathrm GeV^2},\ c=0.400,\ c'=0.010\ \mathrm{GeV}^{-1}\]

Here, the quark mass redefinition  according to
 $2m'_b \equiv 2m_b + F$ was used.\\

The values of the predicted masses, square of the WFOs (or their
derivatives) , and other parameters of interest  for different
bottomonium states obtained using this potential are shown in Table
9. Its it possible to observe  an excellent agreement with experimental
mass values.\\

\newpage

\section{Spin-dependent splittings and $E1$ width: Numerical Results}

In this section 
we show some results from the above mentioned files. 
We would like to stress that such results are not
exhaustive, but just to illustrate how \emph{QQ-onia} runs.

\begin{table}[htb]
\caption{$\Delta M = M(^3S_1)-\ M(^1S_0)$ values (in $\mathrm{MeV}$).}
\label{table:12}
\newcommand{\m}{\hphantom{$-$}}
\newcommand{\cc}[1]{\multicolumn{1}{c}{#1}}
\renewcommand{\tabcolsep}{2pc} 
\renewcommand{\arraystretch}{1.2} 
\begin{tabular}{@{}lllll}
\hline
 $nS$ level & $\Delta M_{Cornell}$ & $\Delta M_{B-T}$ & $\Delta M_{CpL-type}$ & $\Delta M_{Experimental}$ \\
\hline
$1S$ & $135$  & $74$ & $66$ & $71.4$ \\
\hline
$2S$ & $ 57$  & $38$ & $44$ & $----$ \\
\hline
$3S$ & $ 42$  & $28$ & $33$ & $----$ \\
\hline
\end{tabular}\\[2pt]

Parameters employed 
[$|R_{n0}(0)|^2 (\mathrm{GeV}^3)$ and $m_b (\mathrm{GeV})$ ]:\\
 Cornell: $m_b= 5.18 $, $|R_{10}(0)|^2 = 14.06 $, $|R_{20}(0)|^2 = 5.668 $, $|R_{30}(0)|^2 = 4.271$ \\
 B-T: $m_b= 4.88 $, $|R_{10}(0)|^2 = 6.477 $, $|R_{20}(0)|^2 = 3.324 $, $|R_{30}(0)|^2 = 2.474 $\\
 CpL-type: $m_b= 5.1 $, $|R_{10}(0)|^2 = 6.173 $, , $|R_{20}(0)|^2 = 4.027 $, $|R_{30}(0)|^2 = 3.080 $\\
          
\end{table}

\begin{description}

\item[$n^3S_1\ -\ n^1S_0$ splitting:] 
In Table 10 we summarize the results obtained for the  
$n^3S_1\ -\ n^1S_0$ splitting running 
the \emph{SSplit.kumac} file using different potentials. 
According to section \textbf{6}, the parameters involved 
in this calculation are: the heavy-quark mass, its corresponding velocity, and 
the radial wave function at the origin (rWFO). We present our results
from \emph{QQ-onia} using first the set of parameters from the
Cornell potential \cite{eitchten2}. 
We also calculate the splitting employing the parameters from 
the Buchm\"{u}ller and Tye (B-T) QCD-motivated potential \cite{B-T}. 
Lastly, we have reobtained as a final check 
the hyperfine splitting of a Coulomb plus Linear Potential
of Ref.\cite{Patel} (for the particular case $\nu = 1$) .

A comment is in order here. Recently the BaBar collaboration has
claimed the discovery of the long-awaited $\eta_b(1S)$ state \cite{babar}.
Its observed mass ($\simeq 9.389$ GeV) is somewhat lower than expected
yielding a somewhat large hyperfine splitting.   
As can be seen in Table 10, the Cornell potential provides the largest 
deviation (by excess!), obviously due to the fact that this 
potential yields the largest value for the rWFO, while
the rWFO values from the B-T and CpL-type potential are 
considerably smaller. In fact, let us note that
the Buchmuller-Tye model provides an excellent agreement with
the experimental result.

\begin{table}[htb]
\caption{$n(^3P_J)\ and \ n(^1P_1)$ Masses (in GeV); Cornell potential.}
\label{table:13}
\newcommand{\m}{\hphantom{$-$}}
\newcommand{\cc}[1]{\multicolumn{1}{c}{#1}}
\renewcommand{\tabcolsep}{2pc} 
\renewcommand{\arraystretch}{1.2} 
\begin{tabular}{@{}lll}
\hline
 $nS$ level & \emph{QQ-onia} & Experimental\\
\hline
$1^3P_0$ & $9.867$  & $9.860$ \\
$1^3P_1$ & $9.893$  & $9.893$ \\
$1^1P_1$ & $9.900$  & $-----$ \\
$1^3P_2$ & $9.911$  & $9.913$ \\
$2^3P_0$ & $10.236$  & $10.232$ \\
$2^3P_1$ & $10.256$  & $10.255$ \\
$2^1P_1$ & $10.262$  & $-----$ \\
$2^3P_2$ & $10.271$  & $10.269$ \\
\hline
\end{tabular}\\[2pt]
\end{table}

\item[$n^3P_J$ and $n^1P_1$ splitting:] In Table 11 we summarize the 
results obtained from the \newline
\texttt{Split-nP.kumac} files 
using the Cornell potential, in accordance with eqs.(42, 43), to be
compared with the experimental data from \cite{pdg2}.

\item[E1 $2(^3S_1)\rightarrow 1(^3P_J)$ Transitions:] To illustrate this point 
we show in Table 12 the result obtained from the \emph{E1-2S1P.kumac} file, 
in accordance with eq.(44). The Cornell Potential was employed 
to generate the initial and final state wave functions. 
A nice agreement with experimental data is found.

\begin{table}[htb]
\caption{E1 [$2 (^3S_1)\rightarrow 1(^3P_J)$]. From \emph{QQ-onia} $<f\ |\ r\ |\ i> = 1.6915$\ $\mathrm{GeV}^{-1}$. }
\label{table:14}
\newcommand{\m}{\hphantom{$-$}}
\newcommand{\cc}[1]{\multicolumn{1}{c}{#1}}
\renewcommand{\tabcolsep}{0.5pc} 
\renewcommand{\arraystretch}{1.2} 
\begin{tabular}{@{}llll}
\hline
Final state & Photon\ energy (MeV)  & Width (keV)&  Experiment\ (keV)\\
\hline
$J = 0$ & $162.48$ & $1.47$ & $1.22\pm 0.16$ \\
\hline
$J = 1$ & $129.63$ & $2.25$ & $2.21\pm 0.22$ \\
\hline
$J = 2$ & $110.44$ & $2.32$ & $2.29\pm 0.22$ \\
\hline
\end{tabular}\\[2pt]
Experimental data obtained from \cite{pdg2}.
\end{table}

\end{description}

\section{Summary}

The main goal of this work is to
provide a multipurpose (user-friendly) package to obtain the wave functions at
the origin and other relevant properties of heavy-quarkonium
systems, assuming a basic knowledge of the \emph{PAW} software by
the interested reader.\\

Besides, we would like to stress some special peculiarities of our
package, such as providing an easy procedure to normalize the
resulting wave functions on account of the the Numerov
forward-backward framework. In addition, the calculations of the
$l$-derivatives at the origin for angular momentum $l = 2,3,...$ only
require a numerical computation of the wave function and first
derivative through the analytic expressions given in the main text,
thereby keeping the suitable precision for higher derivatives .\\

In addition, we present an alternative method to estimate the
heavy-quark velocity using the well-known virial theorem and a
quick $6^{th}$-order integration, which can be considered as well as a check
of the traditional calculation. In fact, this method can also be interpreted
as an indirect test of the goodness of the potential-probe together with the
(whole range) prior calculated wave function,
since it uses the expectation value of
$r$ times the derivative of the potential $\langle\ r\ V' (r)\ \rangle$.\\

Moreover, some worked examples are presented for the
 bottomonium system using a Cornell-type potential. 
Finally, an additional machinery has been implemented in the code
containing files to analyze the impact of the spin-dependent terms 
in the potential, as well as a tool dealing with E1 transitions. 
Another set of worked examples is presented in this regard.

\subsection*{Acknowledgments}
J.L.D.G gratefully acknowledges his Department for financial support,
and M.A.S.L. thanks MICINN and Generalitat Valenciana 
for financial support under grants FPA2005-01678,
FPA2008-02878 and GVPRE/2008/003. \\

\end{document}